# Charge, geometry, and effective mass in the Kerr-Newman solution to the Einstein field equations

Gerald E. Marsh

Argonne National Laboratory (Ret)
5433 East View Park
Chicago, IL 60615

E-mail: gemarsh@uchicago.edu

**Abstract.** It has been shown that for the Reissner-Nordström solution to the vacuum Einstein field equations charge, like mass, has a unique space-time signature [*Found. Phys*. **38**, 293-300 (2008)]. The presence of charge results in a negative curvature. This work, which includes a discussion of effective mass, is extended here to the Kerr-Newman solution.

**Introduction.**

It has been shown in the predecessor to this paper [1] that if the source of the field is the singularity of the vacuum Reissner-Nordström solution of the coupled Einstein-Maxwell field equations, only the Schwarzschild mass is seen at infinity, with the charge and its electric field making no contribution. It was also shown that if the charge alone is the source of the field, the effective mass seen at infinity vanishes. These results are a direct consequence of charge having the properties of a negative mass while the electric field produced by the charge has a positive effective mass. Near the time-like singularity, which is the source of the field, space-time has a negative curvature. The effective mass within a sphere of radius $R$ was found to be

$$M_{Eff}^{In} = m - \frac{Q^2}{R}. \tag{1}$$

On the other hand, the effective mass of the electric field outside this sphere is given by

$$M_{Eff}^{Out} = \frac{Q^2}{R}. \tag{2}$$

This results in

$$M_{Eff}^{In} + M_{Eff}^{Out} = m. \tag{3}$$

Thus, the "negative mass" associated with the charge $Q$ is exactly compensated by the effective mass contained in the electric field present in the volume exterior to the surface $r = R$. If the radius $r = R$, the effective mass contained within the surface at infinity is $m$, the Schwarzschild or, equivalently, the ADM mass.

These results are extended here to the Kerr-Newman solution. Because of the axial rather than spherical symmetry, the situation is far more complex than in the case of the Reissner-Nordström solution. The Kerr-Newman solution to the field equations is, of course, the charged Kerr solution, and like the Kerr solution, is asymptotically flat. Being axially symmetric, it has two independent, commuting Killing vectors and these have been used [2, 3, 4] in Komar's formula [5] to derive conserved quantities.

The next section addresses the question of spatial curvature near the singularity and this is followed by a discussion of effective mass.

**Curvature in the Kerr-Newman solution**

Vacuum solutions to the Einstein field equations satisfy $R_{\mu\nu} = 0$, and this means that the curvature scalar $R = R^{\mu}{}_{\mu}$ also vanishes. As a consequence, how best to characterize the curvature near black holes is unclear. The Kretschmann scalar $K = R_{\mu\nu\rho\sigma} R^{\mu\nu\rho\sigma}$ has been used by several authors [6, 7], but the interpretation of the scalar is somewhat ambiguous. As pointed out in [7], the Kretschmann scalar can be positive for spaces having a negative curvature. A different method of characterizing the curvature that was used in [1] will be explored below.

The Kerr-Newman solution in generalized Eddington coordinates [8], which are convenient for this approach, is given by

$$ds^2 = dr^2 - 2a \sin^2\theta \, dr \, d\phi + (r^2 + a^2) \sin^2\theta \, d\phi^2 + (r^2 + a^2 \cos^2\theta) d\theta^2 - dt^2 + \frac{2mr - Q^2}{r^2 + a^2 \cos^2\theta} [dr - a \sin^2\theta \, d\phi + dt]^2,$$

(4)

where the symbols have their conventional meanings.

In [1] it was pointed out that for the Reissner-Nordström solution the metric takes the Minkowski form when $r = Q^2/2m$. Interestingly enough, the same thing occurs in the Kerr-Newman metric except that now $r$ has a different meaning with surfaces of constant $r$ corresponding to confocal ellipsoids satisfying

$$\frac{x^2 + y^2}{r^2 + a^2} + \frac{z^2}{r^2} = 1.$$

(5)

It will be seen, however, that unlike the Reissner-Nordström solution, where it was possible to show that for $r < Q^2/2m$ the curvature was negative, the case of the Kerr-Newman solution is more complex.

In order to interpret what follows, it will be necessary to compute the surfaces where $g_{rr}$ and $g_{\theta\theta}$ vanish; the infinite red shift surfaces (where $g_{00} = 0$) for various values of the parameters, and their relation to the horizons, have been given elsewhere [9] and are not relevant to the following discussion. To determine the curvature near the ring singularity, it is useful to examine the case where $a^2 + Q^2 > m^2$, which allows the region near the singularity to be visible from infinity. Setting $a = m = Q = 1$ as a convenient choice of parameters, the condition $g_{\theta\theta} = 0$ results in the fourth order equation

$$r^4 + r^2 + (r^2 + 1)\cos^2\theta + (2r - 1)\sin^2\theta = 0. \tag{6}$$

Two of the solutions to this equation are imaginary, one is negative, and the last is positive and real. The latter is very long and writing it out would add no insight into its nature. The condition $g_{rr} = 0$ using the same value for the parameters results in the quadratic equation

$$r^2 + 2r + \cos^2\theta - 1 = 0. \tag{7}$$

The positive root is $r = \sqrt{2 - \cos^2\theta} - 1$. While the surface $g_{rr} = 0$ will play no role in what follows, it and its relation to the surface $g_{\theta\theta} = 0$ are of interest in their own right. The plots of the real, positive solution to Eq. (6) and of Eq. (7) are shown in Cartesian coordinates in Fig. 1, where henceforth $R = (x^2 + y^2)^{1/2}$. $g_{\theta\theta}$ and $g_{rr}$ are negative (time-like) within their respective toroids $g_{\theta\theta} = 0$ and $g_{rr} = 0$.

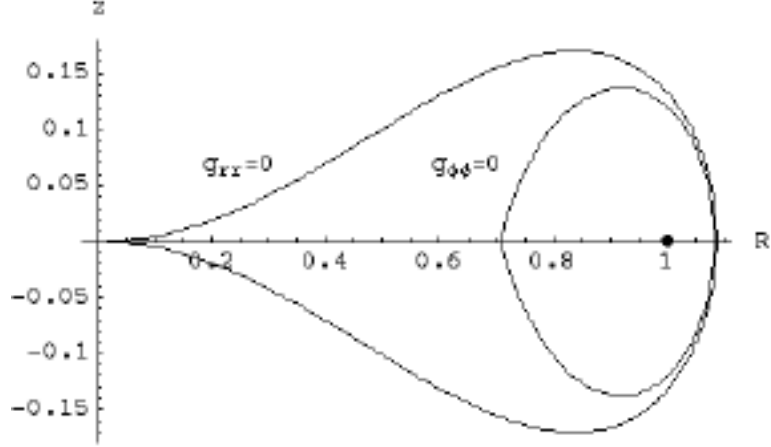

Figure 1. The surfaces where $g_{rr} = 0$ and $g_{\phi\phi} = 0$ in Cartesian coordinates. The values of the parameters are $a = m = Q = 1$. The surfaces are toroids, the $z$-axis being the axis of symmetry. The ring singularity is shown as a heavy dot located at $R = 1$ and $z = 0$.

The method of exploring the curvature near the ring singularity is the same as that used in [1] where the ratio of the circumference of a circle to its radius was computed. One has, for the Eddington coordinates used in Eq. (4), the relations

$$R = (r^2 + a^2)^{\frac{1}{2}} \sin\theta \quad \text{and} \quad z = r \cos\theta. \tag{8}$$

Notice that substituting these equations into Eq. (5) yields an identity.

First consider the equatorial plane where $\theta = \pi/2$. The first of Eqs (8) allow the ratio of the circumference of a circle in the plane to its radius to be written as

$$\frac{C}{R} = \frac{\int_0^{2\pi} \sqrt{g_{\phi\phi}}\, d\phi}{(r^2 + a^2)^{\frac{1}{2}}} = 2\pi \left( \frac{\dfrac{(2mr - Q^2)a^2}{r^2} + r^2 + a^2}{(r^2 + a^2)} \right)^{\frac{1}{2}}. \tag{9}$$

Note that when $C/R$ is set equal to $2\pi$ and the resulting equation solved for $r$ one obtains $r = Q^2/2m$ where the metric of Eq. (4) takes the Minkowski form. This ratio is plotted in Fig. (2) for the same values of the parameters used in Fig. (1). The ratio vanishes for $r \sim 0.404698$, which corresponds to $R \sim 1.0830$ in Cartesian coordinates. The toroidal surface $g = 0$ of Fig. 1 intersects the equatorial plane in two circles. The latter values of $r$ or $R$ correspond to the circle of greatest radius.

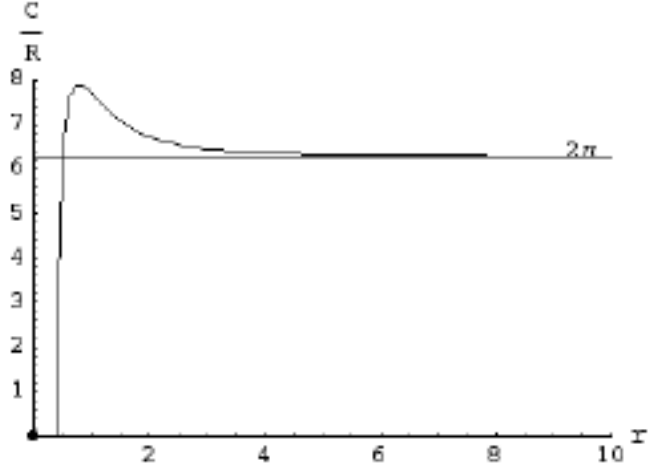

Figure 2. The ratio of the circumference to the radius $R$ in the equatorial plane of the Kerr-Newman solution in Eddington coordinates. The ring singularity is shown as the heavy dot at $r = 0$. The portion of the curve above the line $C/R = 2$ corresponds to a negative curvature and that below to positive curvature. $C/R = 0$ at $r \sim 0.404698$ where $g = 0$, and crosses the line $C/R = 2$ at $r = Q^2/2m$, which for the value of the parameters used here, $a = m = Q = 1$, is 0.5.

The ratio of the circumference to its radius for a circle not in the equatorial plane is considerably more complicated. Using both of Eqs. (8), the ratio can be written as a function of the variables and $z$ as

$$\frac{C}{R} = \frac{\int_0^{2\pi} \sqrt{g}\, d\phi}{(r^2 + a^2)^{\frac{1}{2}} \sin} =$$

$$\frac{2\pi \left( \frac{(2mz\cos - Q^2\cos^2)a^2\sin^4}{z^2 + \cos^4} + (z^2 + a^2\cos^2)\tan^2 \right)^{\frac{1}{2}}}{(z^2 + a^2\cos^2)^{\frac{1}{2}}\tan}.$$

(10)

For a given plane $z =$ Const., each value of $\theta$ determines a circle in the plane centered on the z-axis, corresponding to the intersection of that plane with the hyperboloid of one sheet associated with each $\theta$. These hyperbolae are confocal to the ellipsoids corresponding to $r =$ Const. given by Eq. (5). There is a bit of a subtlety here in that the surface $\theta =$ Const. is only a half-hyperboloid lying in the half space $z > 0$ ($z < 0$) when $\theta < \pi/2$ ($\theta > \pi/2$).

Note again that if $z = r \cos\theta$ is substituted back into the expression for $C/R$ given by Eq. (10), and the result set equal to $2\pi$, the solution to the equation for $r$ again yields $r = Q^2/2m$ independent of $\theta$. The plots for various values of $z$ are shown in Figs. 3(a), 3(b), and 3(c).

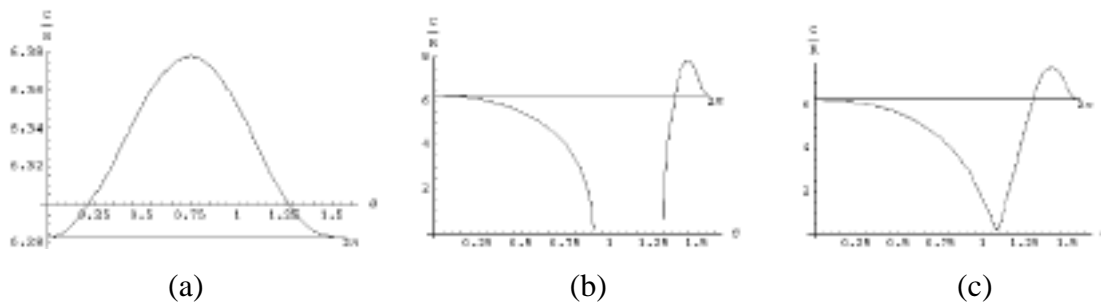

(a)                                    (b)                                  (c)

Figure 3. The ratio of the circumference to the radius for three values of $z$. Note that each value of $\theta$ corresponds to a different circle centered on the z-axis in the plane $z =$ Const. (a) $z = 2$; (b) $z = 0.1$. The ratio vanishes at $\theta \approx 0.905169$ and $\theta \approx 1.29788$ where the plane $z = 0.1$ intersects the toroid $g_{tt} = 0$; (c) $z \approx 0.13569$. Here the plane is almost tangent to the toroidal surface of $g_{tt} = 0$. In these figures, $\theta = 0$ corresponds to the $C/R$-axis and $\theta = \pi/2$ to infinity. The curves in (b) and (c) cross the line $C/R = 2\pi$ at $r = Q^2/2m = 0.5$ for the chosen parameters.

Interpretation of Fig. 3, compared to that of the Reissner-Nordström solution, is complicated by the effects of rotation. The difference in behavior between the Kerr and Kerr-Newman solution is due to there being no toroidal surface within which $g_{tt}$ is time-like for the uncharged Kerr metric. An excellent discussion of rotationally induced effects in the Kerr metric has been given by de Felice and Clarke [10].

The effects of rotation in the presence of a magnetic field have been studied by Kulkarni and Dadhich [11], who also discuss the Gaussian curvature and its role in the embedding problem.

Thus, while the computation of the ratio of the circumference of a circle to its radius in a plane $z = $ constant gives some interesting results, perhaps the most important conclusion is that for $r = Q^2/2m$ the ratio is equal to 2 . It is at this radius that the Kerr-Newman metric takes the Minkowski form.

**Effective mass in the Kerr-Newman solution**

From Eq. (1) it can be seen that in the case of the Reissner-Nordström solution the effective mass within a sphere of radius $r = Q^2/2m$ is $-m$. While the literature contains a number of definitions for the Kerr-Newman effective mass [2, 4], the one that will be used here is that given by de Felice and Bradley [6]. The latter not only has the virtue of being equal to the Schwarzschild or ADM mass at infinity, it also yields the Reissner-Nordström value of $-m$ at the radius $r = Q^2/2m$. de Felice and Bradley give the equation

$$M_{KN} = m - \frac{2m\,a^2\cos^2\theta + Q^2 r}{r^2 + a^2\cos^2\theta} .$$
(11)

This expression for the Kerr-Newman effective mass is dependent on the variable $\theta$, but for $r = Q^2/2m$ it yields $-m$ independent of $\theta$. This means that the region within the surface $r = Q^2/2m$ effectively has negative curvature. Since the value of the effective mass at infinity is $m$, the effective mass of the field energy contained in region $Q^2/2m \leq r \leq \infty$ must be $2m$. Again the same as the Reissner-Nordström solution despite, as discussed earlier, the difference in meaning for $r$.

As is readily apparent, if $Q^2 = 0$ the effective mass given by Eq. (11) vanishes on the surface $r = a\cos\theta$. Indeed, the origin of Eq. (11) is related to the fact that the Kretschmann scalar vanishes on this surface for the uncharged Kerr metric. The Kretschmann scalar for the Kerr-Newman metric is given by [7]

$$K = \frac{8}{(r^2 + a^2\cos^2\theta)^6}$$
$$\times \Big[ 6m^2(r^6 - 15\,a^2 r^4\cos^2\theta + 15\,a^4 r^2\cos^4\theta - a^6\cos^6\theta) - 12\,m\,r\,Q^2$$
$$(r^4 - 10\,a^2 r^2\cos^2\theta + 5\,a^4\cos^4\theta) + Q^4(7\,r^4 - 34\,a^2 r^2\cos^2\theta + 7\,a^4\cos^4\theta)\Big].$$
(12)

For $Q^2 = 0$ and $r = a \cos\theta$ one may confirm that $K = 0$. However, this is not the case for $Q^2 \neq 0$, nor is it true for the surface $r = Q^2/2m$. It is therefore not clear what role the Kretschmann scalar plays for the Kerr-Newman solution.

Cohen and de Felice, in an earlier paper [2], derive an expression for the effective mass of the Kerr-Newman solution in Boyer-Lindquist coordinates. The metric in these coordinates is

$$ds^2 = -\left(\frac{\Delta}{A}\right)dt^2 + \left(\frac{A \sin^2\theta}{\Sigma}\right)[d\phi + \omega dt]^2 + \left(\frac{\Sigma}{\Delta}\right)dr^2 + \Sigma d\theta^2,$$

(13)

where,

$$\omega = \frac{a}{A}\left(Q^2 - 2mr\right)$$
$$\Delta = r^2 + a^2 + Q^2 - 2mr$$
$$\Sigma = r^2 + a^2 \cos^2\theta$$
$$A = \left(r^2 + a^2\right)\Sigma - a^2\left(Q^2 - 2mr\right)\sin^2\theta.$$

(14)

Cohen and de Felice then evaluate the Komar integral [5]

$$I = \oint_{\partial V} *d\xi,$$

(15)

where $\xi$, is the Killing 1-form, $*d\xi$ is the Hodge dual of the 2-form $d\xi$, and $V$ is the volume interior to the space-like boundary $\partial V$. They then use an orthonormal frame of 1-forms to evaluate $*d\xi$. The surface chosen is $r = r_0 = $ constant. Because of the cross terms in the metric, the "time" difference for simultaneous events [12] on this surface differ by $dt = -g_{00}^{-1} g_{0\phi} d\phi$. So as to have a surface of simultaneous events one must subtract the contribution to the integral due to the time difference between initial and final events. The Komar integral then reduces to what will be called the effective mass interior to the surface $r = r_0$:

$$M^{Int}_{K-N} = \frac{1}{8\pi}\oint_V *d\xi = \frac{1}{8\pi}\int_0^{2\pi} d\phi \int_0^\pi \tilde{f}A^{\frac{1}{2}}\sin\theta\, d\theta,$$

(16)

where (including a sign correction from [6]),

$$\tilde{f} = \frac{2A^{\frac{1}{2}}}{\Sigma^3}\left[m\Sigma + (Q^2 - 2mr)r\right]\left(1 + a^2\sin^2\theta\right).$$

(17)

The integral is readily evaluated and yields

$$M^{Int}_{K-N} = m - \frac{Q^2}{2r_0} - \frac{Q^2(r_0^2 + a^2)}{2ar_0^2}\tan^{-1}\left(\frac{a}{r_0}\right).$$

(18)

As pointed out by Cohen and de Felice, this expression does not explicitly include the negative contribution to the effective mass due to rotation. This is readily apparent by setting $Q^2 = 0$, which leaves only the mass term. Kulkarni, et al. and Chellathurai and Dadhich [3, 4] give an expression for the effective mass that does include rotation, however it is exact only in the limits of the outer horizon and infinity. Since the interest here is primarily on the effects of charge, the expression given by Eq. (18) suffices.

That Eq. (18) yields the correct value for the Reissner-Nordström solution when $a = 0$ can be seen by expanding the right hand side in a series in $a$. One obtains

$$M^{Int}_{K-N} = m - \frac{Q^2}{r_0} - \frac{Q^2 a^2}{3r_0^3} + \frac{Q^2 a^4}{15r_0^5} - \frac{Q^2 a^6}{35r_0^7} + \ldots.$$

(19)

For $a = 0$, the result is the same as in Eq. (1).

Cohen and de Felice also derive the effective mass due to the electric field in a volume *interior* to $r = r_0$. The approach they used also allows one to find the effective mass of the electric field in the volume $r_0 \leq r \leq \infty$ *exterior* to $r_0$. To do this one uses the contour shown in Fig. 4.

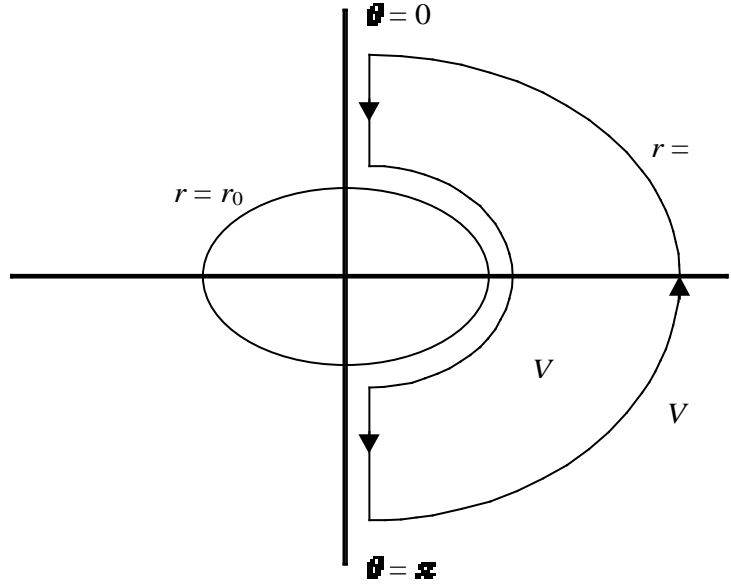

Figure 4. The integration of Eq. (20) is over the closed toroidal surface of the volume *V*.

The Komar integral for the contour of Fig. 4 is given by

$$M_{K-N}^{Ext} = \frac{1}{8\pi} \oint_V *d\xi =$$

$$\frac{1}{4} \left[ \int_0^{\pi/2} \tilde{f} A^{\frac{1}{2}} \sin\theta \, d\theta \Big|_{r=r_0^+} + \int_\infty^0 \tilde{f} A^{\frac{1}{2}} \sin\theta \, d\theta \Big|_{r=\infty} \right]$$

$$-\frac{1}{4} \left[ \int_\infty^{r_0} \tilde{g} \left(\frac{A}{\Delta}\right)^{\frac{1}{2}} \sin\theta \, dr \Big|_{\theta=0} + \int_{r_0}^\infty \tilde{g} \left(\frac{A}{\Delta}\right)^{\frac{1}{2}} \sin\theta \, dr \Big|_{\theta=\pi} \right],$$

(20)

where,

$$\tilde{g} = a \left(\frac{A^3}{6}\right)^{\frac{1}{2}} \sin 2 \left[1 + \left(\frac{r^2 + a^2}{a}\right)\right].$$

(21)

In doing the integration, one must also take into account the path of integration and the fact that for the volume $V$ the normal to the surface $r = r_0$ points in the negative $r$-direction (opposite to the direction when computing the effective mass within this surface). The last two integrals in Eq. (20) vanish, and the first two combine to both eliminate the mass term associated with the ring singularity (which is exterior to the volume of integration) and yield

$$M_{\text{K-N}}^{Ext} = \frac{Q^2}{2r_0} + \frac{Q^2(r_0^2 + a^2)}{2ar_0^2} \tan^{-1}\left(\frac{a}{r_0}\right).$$

(22)

Thus, as in the case for the Reissner-Nordström solution, one has

$$M_{\text{K-N}}^{Int} + M_{\text{K-N}}^{Ext} = m,$$

(23)

where $m$ is the Schwarzschild or ADM mass.

**Summary**

By excluding the effects of rotation in the definition of effective mass for the Kerr-Newman metric, it has been shown that the "negative mass" due to charge has properties very similar to that of the Reissner-Nordström metric. Both take the Minkowski form at $r = Q^2/2m$, even though the meaning of $r$ is different for the two metrics; the effective mass interior to this surface is $-m$ in both cases; and both have an effective mass of *m* at infinity. In addition, the effective mass for both metrics satisfies, for any surface defined by *r* = constant (again for either definition of *r*), the relation

$$M_{Eff}^{Int} + M_{Eff}^{Ext} = m.$$

(24)

Thus the positive effective mass of the electric field exterior to the surface exactly compensates for the "negative mass" associated with the charge located within the surface.

**Acknowledgement**

The author would like to thank one of the referees for catching two typographical errors and pointing out an interesting and relevant paper.